\documentclass{aastex63}

\usepackage{amsmath,amssymb}
\usepackage{natbib}
\usepackage{lineno}

\usepackage[caption=false]{subfig}
\usepackage{graphicx}
\usepackage{epstopdf}

\usepackage{color}
\usepackage{multirow}
\usepackage{graphicx}
\usepackage{booktabs}
\usepackage[T1]{fontenc}

\graphicspath{{./}{figures/}}

\begin{document}

\title{Hard TeV Gamma-Ray Afterglows of Nearby GRB 190829A as a Tentative Signature of Ultra-High-Energy Cosmic Rays Accelerated in Gamma-Ray Burst Jets}
\correspondingauthor{Xiao-Li Huang, En-Wei Liang}
\email{xiaoli.huang@gznu.edu.cn, lew@gxu.edu.cn}

\author[0000-0002-2542-3873]{Jian-Kun Huang}
\affiliation{Guangxi Key Laboratory for Relativistic Astrophysics, School of Physical Science and Technology, Guangxi University, Nanning 530004, China}

\author[0000-0002-9725-7114]{Xiao-Li Huang$^{*}$}
\affil{School of Physics and Electronic Science, Guizhou Normal University, Guiyang 550025, China}

\author[0000-0002-2585-442X]{Ji-Gui Cheng}
\affiliation{Guangxi Key Laboratory for Relativistic Astrophysics, School of Physical Science and Technology, Guangxi University, Nanning 530004, China}

\author[0000-0002-9037-8642]{Jia Ren}
\affil{School of Astronomy and Space Science, Nanjing University, Nanjing 210023, China}
\affil{Key Laboratory of Modern Astronomy and Astrophysics (Nanjing University), Ministry of Education, Nanjing 210023, China}

\author[0000-0003-0726-7579]{Lu-Lu Zhang}
\affiliation{Guangxi Key Laboratory for Relativistic Astrophysics, School of Physical Science and Technology, Guangxi University, Nanning 530004, China}

\author[0000-0002-7044-733X]{En-Wei Liang$^{*}$}
\affiliation{Guangxi Key Laboratory for Relativistic Astrophysics, School of Physical Science and Technology, Guangxi University, Nanning 530004, China}

\begin{abstract}
The observed hard TeV gamma-ray spectrum of the nearby gamma-ray burst (GRB) 190829A may challenge the conventional leptonic GRB afterglow model.
It has been proposed that an ultra-high-energy (UHE; $\varepsilon^{'}_{\rm p}\sim 10^{20}$ eV) proton population can be pre-accelerated by internal shocks in GRB jets.
We study possible signatures of the UHE protons embedded in the TeV afterglows when they escape the afterglow fireball.
We show that the leptonic model can represent the observed multiwavelength lightcurves and spectral energy distributions of GRB 190829A by considering the uncertainties of the model parameters.
Attributing the TeV gamma-ray afterglows to the emission of both the electron self-Compton scattering process and the UHE proton synchrotron radiations in the afterglow fireball, we obtain tentative upper limits of $\log_{10} \varepsilon_{\rm p}^{\prime}/{\rm eV}\sim 20.46$ and $\log_{10}E_{\rm p, total}/{\rm erg}\leq 50.75$, where $E_{\rm p, total}$ is the total energy of the proton population.
The synchrotron radiations of the UHE protons should dominate the early TeV gamma-ray afterglows, implying that early observations are critical for revealing the UHE proton population.
\end{abstract}


\section{Introduction}
\label{sec:intro}
Gamma-ray bursts (GRBs) are bright gamma-ray flashes from ultrarelativistic jets powered by collapses of massive stars or mergers of compact objects \citep{2004RvMP...76.1143P, 2015PhR...561....1K}.
In the framework of the standard internal and external fireball models, both the prompt gamma rays and multiwavelength afterglows are attributed to the emission of electrons accelerated in jets via synchrotron radiation and synchrotron self-Compton (SSC) processes \citep{1993ApJ...418L..59M, 1994ApJ...432..181M, 1997ApJ...482L..29M, 1998ApJ...497L..17S, 2001ApJ...548..787S}.
It has also been speculated that GRB jets are potential accelerators of ultra-high-energy cosmic rays (UHECRs; i.e., $\varepsilon_{\rm p}>10^{19}$ eV; \citealp{1995ApJ...449L..37M, 1995ApJ...453..883V,1995PhRvL..75..386W,2011MNRAS.418.1382L}).
Observations of very-high-energy (VHE; $\sim$ TeV) gamma-ray photons provide a new way to study the particle acceleration and radiation physics of GRB jets (e.g. \citealp{2019Natur.575..448Z}).
Significant progress on observations of VHE gamma-ray afterglows has been made with current ground-based telescopes.
The Major Atmospheric Gamma Imaging Cherenkov(MAGIC) telescopes convincingly detected the sub-TeV gamma-ray emission in the early afterglow of GRB 190114C at a redshift of 0.4245 \citep{2019Natur.575..455M}.
Its broadband spectral energy distributions (SEDs) can be explained with the leptonic model, which interprets the optical X-ray emission as electron synchrotron radiations and attributes the GeV-TeV gamma rays to the SSC process \citep{2019ApJ...880L..27D,2019Natur.575..459M,2019ApJ...884..117W}.
Actually, signatures of the SSC components are also seen in GRB 130427A \citep{2013ApJ...773L..20L,2014Sci...343...42A,2020ApJ...903L..26H} and GRB 180720B \citep{2019ApJ...884...61D,2019ApJ...885...29F,2019ApJ...884..117W}.
More interestingly, the High Energy Stereoscopic System (H.E.S.S.) detected VHE gamma-ray photons ($\sim 4$ TeV) of nearby GRB 190829A at $z = 0.0785 \pm 0.005$ \citep{2021Sci...372.1081H}.
The VHE gamma-ray emission was detected at a very late epoch, i.e. at $t\sim 2 \times 10^{4}$ s and even up to $t=2\times 10^{5}$ s after the GRB trigger observed by the Gamma-Ray Burst Monitor on board the Fermi Gamma-Ray Space Telescope.
GRB 190829A shares similar properties with low-luminosity GRBs (LL-GRBs), such as low luminosity, low Lorentz factor, and large jet opening angle \citep{2006ApJ...646..351L}.
The detection of VHE gamma rays from GRB 190829A indicates that LL-GRBs are also VHE gamma-ray sources \citep{2020ApJ...898...42C,2021ApJ...918...12F} and potential UHECR sources \citep{2006PhRvD..73f3002M,2011MNRAS.418.1382L}.
As a nearby source, its VHE gamma-ray photons suffer less absorption by the extragalactic background light (EBL; \cite{1966PhRvL..16..252G}).
The VHE data, together with abundant radio-optical X-ray afterglow data \citep{2020MNRAS.496.3326R, 2021A&A...646A..50H, 2022ApJ...931L..19S}, make GRB 190829A a valuable case for investigating particle acceleration and radiative physics of GRB jets.

The radio-optical X-ray afterglow data of GRB 190829A can be explained as the emission from the synchrotron radiations of the electrons accelerated in the jet \citep{2021Sci...372.1081H,2021ApJ...917...95Z}.
However, the radiation mechanism of the VHE gamma-ray emission of GRB 190829A is being debated.
As reported by \cite{2021Sci...372.1081H}, the power-law spectral index of the gamma rays between 0.18 and 3.3 TeV is very hard ($\beta_\gamma=2.07\pm0.09$) and the SSC process of the electrons cannot properly explain the hard spectrum (see also \citealp{ 2022ApJ...925..182H}).
However, it has also been suggested that the SSC process or the external inverse-Compton scattering process can yield the VHE gamma-ray flux at $t=10^4$ s after the GRB trigger (e.g., \citealp{2021ApJ...920...55Z}).
It has also been proposed that a hadronic process could be involved in explaining the hard spectrum component \citep{2022ApJ...929...70S}.

Analytical and simulation analyses show that GRBs are potential acceleration sites of UHECRs (e.g. \citealp{1995PhRvL..75..386W,2021ApJ...908..193M,2022MNRAS.511.5823R}).
The magnetized shocks could accelerate protons up to $10^{21}$ eV \citep{1987PhR...154....1B,1993A&A...272..161R}.
The most conventional acceleration mechanism of UHECRs is the Fermi acceleration of protons at astrophysical magnetized shocks in relativistic jets of GRBs and active galactic nuclei.
The accelerated region of shocks could be regarded as a leaky box \citep{1999APh....10..185P}.
Although some protons may escape the leaky box, a small fraction of protons could be confined and repeatedly accelerated to the UHE energy band and piled up at a certain energy, which depends on the physical condition of the shocks \citep{1997PASA...14..251M,1999A&A...347..370D,1999APh....10..185P}.
The protons are continuously accelerated and cooled.
Simulation analysis shows that a tiny fraction of the protons could be exponentially accelerated to high energy but saturated at $\varepsilon_{\rm p}^{\prime}\sim 10^{20}$ eV in a magnetic field $B\sim 10$ G \citep{2018PhRvL.121x5101A,2021ApJ...908..193M,2022MNRAS.511.5823R}, shaping the proton energy spectrum as a cutoff power-law function or a cutoff power-law function plus a high-energy bump/peak, which have been proposed to explain the unusual TeV gamma-ray spectra of Mrk 421 and Mrk 501 \citep{2000NewA....5..377A}.
In addition, stochastic acceleration can also result in the formation of a pileup of electrons \citep{2005ApJ...621..313V, 2008ApJ...681.1725S, 2011ApJ...739...66T}, which has been adopted to explain the sharp and narrow spectral feature observed in the VHE band of Mrk 501 \citep{2020A&A...637A..86M}.

TeV gamma rays are important messengers of UHE protons \citep{1998ApJ...502L..13T,1998ApJ...509L..81T}.
These UHE protons emit TeV gamma rays via the synchrotron radiations.
The detection of VHE gamma rays ($\sim 4$ TeV) from GRB 190829A may be evidence of the UHE protons accelerated in the GRB jets.
This paper investigates this issue with the late VHE gamma-ray afterglow data of GRB 190829A.
We organize our paper as follows.
We describe our model in \S 2 and present our numerical results in \S 3. Discussion and conclusions are available in \S 4 and \S 5.

\section{Models}
\label{sec:Model}
\subsection{UHE Protons and Their Synchrotron Radiations}
As discussed by \cite{1999APh....10..185P}, charged particles could be accelerated to the saturation energy if their energy-loss timescale ($t_{\rm loss}$) is equal to their acceleration timescale ($t_{\rm acc}$), shaping the spectrum of the accelerated particles as a cutoff power-law function or that with a high-energy bump/peak \citep{2000NewA....5..377A}.
The \cite{2020A&A...637A..86M} described such a spectrum with a relativistic Maxwellian distribution.
It has been suggested that a fraction of the protons in the GRB jet could be exponentially accelerated to high energy but saturated at an energy up to $\sim 10^{20}$ eV \citep{2018PhRvL.121x5101A,2021ApJ...908..193M,2022MNRAS.511.5823R}.
For simplicity, we adopt a monochromatic distribution, $N_{\rm p}^{\prime}(\varepsilon_{\rm p}^{\prime})=N_{\rm p,0}^{\prime}\delta(\varepsilon_{\rm p}^{\prime}-\varepsilon_{\rm p,0}^{\prime})$, to approximate the distribution of the {\bf saturation} UHE protons, where "$\prime$" marks the comoving frame of shocks.
This is comparable to the extreme scenario of a spiky proton distribution presented by \cite{2000NewA....5..377A}.
Further discussions on the UHE protons and electrons in the prompt gamma rays are available in \S 4.1.

The UHE proton population is injected into the afterglow jet of GRB 190829A.
These protons experience adiabatic expansion, if one does not consider the acceleration by the external shocks and the energy dissipation of the protons.
They emit TeV gamma rays via the synchrotron radiation mechanism in the external shock environment.
The synchrotron emissivity of the protons in the comoving frame is calculated as \citep{1986A&A...164L..16C}
\begin{equation}
\varepsilon_{\gamma}^{\prime}J_{\rm syn}^{\prime}(\varepsilon_{\gamma}^{\prime})=\frac{\sqrt{3}\varepsilon_{\gamma}^{\prime}e^{3}B^{\prime}}{hm_{\rm p}c^{2}}\int^{\infty}_{0}d\varepsilon_{\rm p}^{\prime}N_{\rm p}^{\prime}(\varepsilon_{\rm p}^{\prime})V^{\prime}R(x),
\end{equation}
where $\varepsilon_{\gamma}^{\prime}$ is the photon energy in the comoving frame, $V^{\prime}$ is the comoving volume of the afterglow radiating region,
$x=\frac{4\pi\varepsilon_{\gamma}^{\prime}m_{\rm p}^{3}c^{5}}{3eB^{\prime}hE_{\rm p}^{\prime,2}}$, and $R(x)$ is given by
\begin{equation}
R(x)=\frac{x}{2}\int^{\pi}_{0}d\theta \sin\theta\int^{\infty}_{x/\sin\theta}dtK_{5/3}(t),
\end{equation}
in which $K_{5/3}(t)$ is the modified Bessel function of the second kind of order $5/3$.
The function $R(x)$ can be approximated with a simple analytical form, as presented in \cite{2008ApJ...686..181F}.
The synchrotron flux in the observer's frame is given by
\begin{equation}
f_{\rm syn}(\varepsilon_{\gamma})=\frac{\delta_{\rm D}^{4}}{4\pi d_{\rm L}^{2}}\varepsilon_{\gamma}^{\prime}J_{\rm syn}^{\prime}(\varepsilon_{\gamma}^{\prime})=\frac{\delta_{\rm D}^{4}}{4\pi d_{\rm L}^{2}}\frac{\sqrt{3}\varepsilon_{\gamma}^{\prime}e^{3}B^{\prime}}{hm_{\rm p}c^{2}}\int^{\infty}_{0}d\varepsilon_{\rm p}^{\prime}N_{\rm p}^{\prime}(\varepsilon_{\rm p}^{\prime})V^{\prime}R(x),
\end{equation}
where $\varepsilon_{\gamma}=\varepsilon_{\gamma}^{'}\delta_{\rm D}/(1+z)$ is the photon energy in the observer's frame and $\delta_{\rm D}$ is the Doppler factor.

\subsection{Synchrotron and SSC Processes of the Electrons}
We take the energy distribution of the injected electrons in the energy range of $\gamma^{\prime}_{\rm e,min} \leqslant \gamma^{\prime}_{\rm e} \leqslant \gamma^{\prime}_{\rm e,max}$ as a single power-law function, i.e. $Q(\gamma^{\prime}_{\rm e})\propto {\gamma^{\prime}_{\rm e}}^{-p}$.
The evolution of the typical synchrotron emission frequency, the cooling frequency, and the peak spectral flux read (e.g. \citealp{1998ApJ...497L..17S})
\begin{eqnarray}
&\nu_m&  =  3.3 \times 10^{12}~{\rm Hz}~
\left(\frac{p-2}{p-1}\right)^2(1+z)^{1/2}\epsilon_{\rm B,-2}^{1/2}
\epsilon_{\rm e,-1}^{2}E_{\rm k,52}^{1/2} t_{\rm d}^{-3/2} \label{num},\\
&\nu_c&  =  6.3 \times 10^{15}~{\rm Hz}~
(1+z)^{-1/2} (1+Y)^{-2} \epsilon_{\rm B,-2}^{-3/2}E_{\rm k,52}^{-1/2} n^{-1}
t_{\rm d}^{-1/2} \label{nuc},\\
&F_{\nu,\max}&  = 1.6~{\rm mJy}~
(1+z)D^{-2}_{28}\epsilon_{\rm B,-2}^{1/2}E_{\rm k,52}n^{1/2}
\label{Fnumax},
\end{eqnarray}
where $t_{\rm d}$ is the observer's time in units of days, $D$ is the luminosity, $E_{\rm k}$ is the kinetic energy of the fireball, $n$ is the number density of the medium, $\epsilon_{\rm e}$ and $\epsilon_{\rm B}$ the electron and magnetic field energy partition fractions of the fireball internal energy, respectively, $Y$ is the IC parameter, and the notation $Q_{\rm n}$ is defined as $Q_{\rm w}=Q/10^w$ in cgs units. We have $Y=[-1+(1+4 \epsilon_{\rm rad}\xi \epsilon_{\rm e} / \epsilon_{\rm B})^{1/2}]/2$, where $\xi \leq 1$ is a correction factor introduced by the Klein-Nishina correction, $\epsilon_{\rm rad}=\min\{1,(\gamma_{\rm e,min}/\gamma_{\rm e,c})^{p-2}\}$ \citep{2001ApJ...548..787S, 2008MNRAS.384.1483F}, and $ \gamma_{\rm e,c}=6 \pi m_{\rm e} c /(\sigma_{\rm T} \Gamma {B^{\prime}}^2 t^{\prime})$, in which $c$ is the speed of light, $m_{\rm e}$ is the electron mass, and $\sigma_{\rm T}$ is the Thomson scattering area.
The emission of the SSC process is calculated based on the electron spectrum and seed photons from the synchrotron radiation.
Note that the SSC effect is significantly suppressed in the Klein-Nishina regime for photons with energy $\nu > \nu_{\rm KN}$, where
\begin{eqnarray}
\nu_{\rm KN} & = & h^{-1} \Gamma m_{\rm e} c^2 \gamma_{\rm e,X}^{-1} (1+z)^{-1}
\nonumber \\
& \simeq & 2.4 \times 10^{15} ~{\rm Hz}~ (1+z)^{-3/4} E_{\rm k,52}^{1/4}
\epsilon_{\rm B,-2}^{1/4} t_{\rm d}^{-3/4} \nu_{18}^{-1/2} ~,
\end{eqnarray}
in which $h$ is Planck's constant.
Analytically, we have $\xi =\min\{1, (\nu_{\rm KN} / \nu_{\rm c})^{(3-p)/2}\}$ for slow cooling and $\xi = \min\{1, (\nu_{\rm KN} / \nu_{\rm m})^{1/2}\}$ for fast cooling, where the factors $(\nu_{\rm KN} /\nu_{\rm c})^{(3-p)/2}$ and $(\nu_{\rm KN} / \nu_{\rm m})^{1/2}$ denote the fractions of the photon energy density that contribute to the SSC process in the X-ray band in the slow and fast-cooling regimes, respectively. We have numerically solved the Klein-Nishina effect and the $\gamma\gamma$ annihilation effect.
The synchrotron self-absorption effect is also considered in the calculations. For the detials of our leptonic model, please refer to  \citet{2020ApJ...901L..26R, 2022arXiv221010673R}, and references therein.

\subsection{Fireball Dynamics}
Following \cite{2021ApJ...917...95Z}, we assume that the external fireball is initially a pre-accelerated $e^{\pm}$-rich medium shell in radius of $R_{\rm s}\leq R \leq R_{\rm e}$, then transits to a normal medium at $R>R_{\rm e}$, i.e. $n=kn_0$ if $R_{\rm s} \leq R \leq R_{\rm e}$ and $n=n_0$ if $R>R_{\rm e}$.
The $e^{\pm}$-rich medium shell is proposed as the result of the interaction between the surrounding medium and the initial hard gamma rays of the first prompt pulse of GRB 190829A.
We assume that the GRB jet is an on-axis top-hat jet with a half-opening angle of $\theta_{\rm j}$.
The jet lateral expansion is not considered.
We divide the radiation surface into grids/patches and calculate the radiations of each patch.
The effect of the equal arrival time surface (EATS) is considered.
We numerically calculate the arrival-time of photons from each patch.
The intrinsic lightcurves and SEDs can be obtained from the integration over EATS after considering the Doppler boosting effect.
For the details of our dynamics model of the external shock fireball please refer to \citet{2020ApJ...901L..26R}.

\section{Data and Numerical Results}
\label{sec:Fit}
\subsection{Data}
\begin{figure}[ht]	
	\centering
	\includegraphics[scale=0.14]{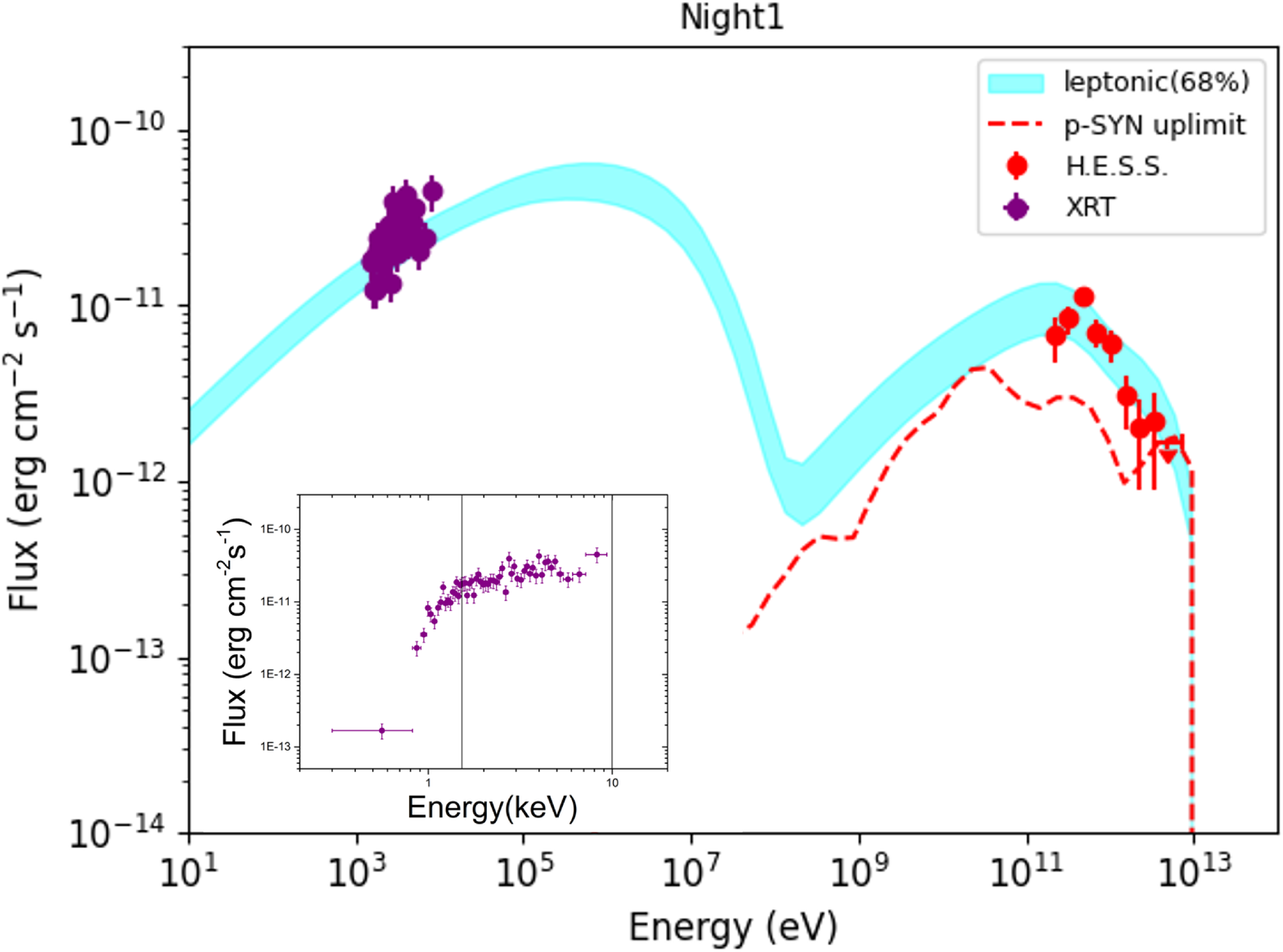}
	\includegraphics[scale=0.14]{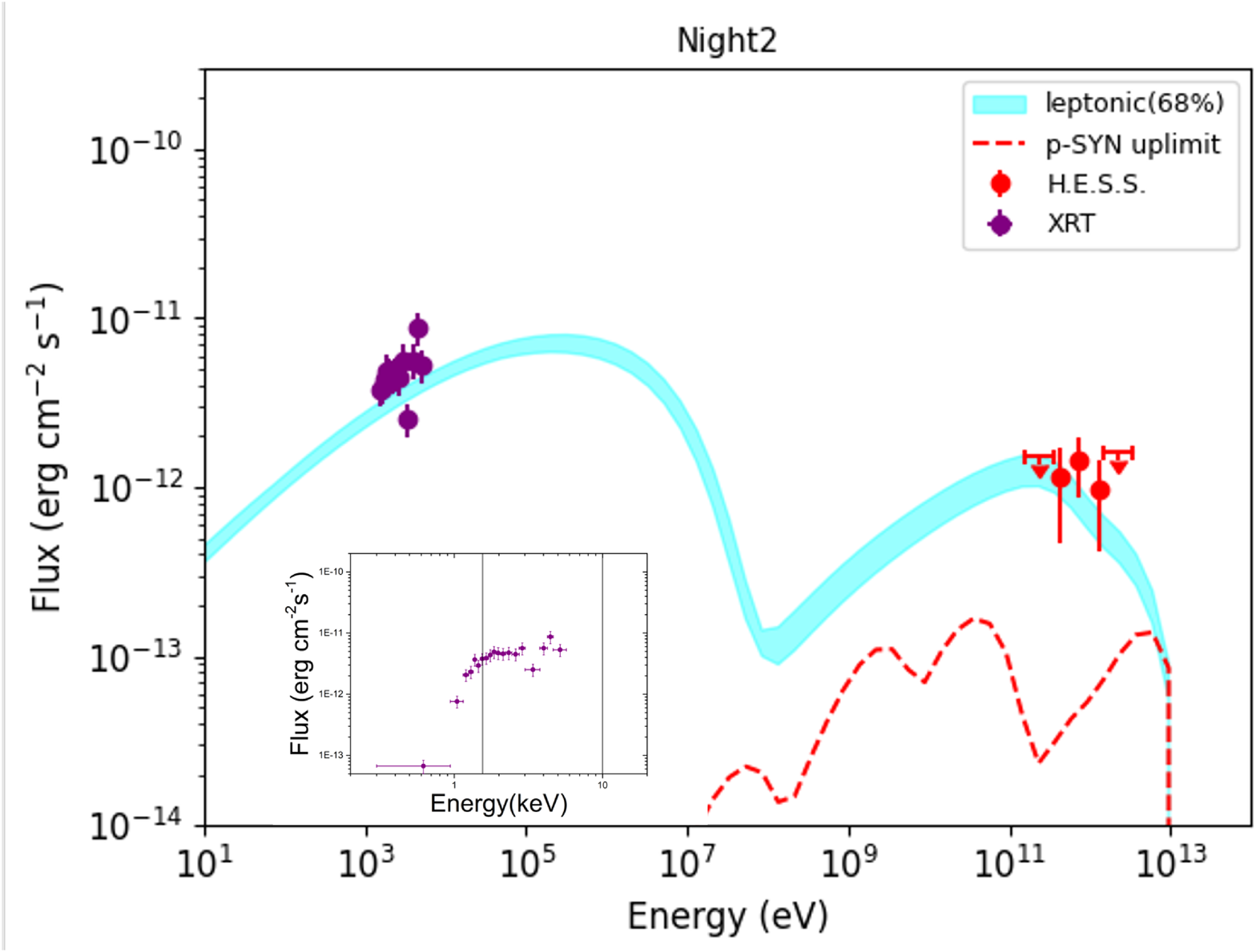}
	\caption{Afterglow SEDs of GRB 190829A observed with H.E.S.S (0.2-4 TeV; data points) and our lepto-hadronic model fit (bands). The insets illustrate the X-ray spectra in detail, where the vertical lines mark the energy band of 1.5-10 keV. The gamma-ray flux is corrected by the EBL absorption.}
	\label{SED}
\end{figure}

The multiwavelength afterglow lightcurves of GRB 190829A are shown in Figure \ref{LC}, and the afterglow SEDs in the X-ray and TeV gamma-ray bands are shown in Figure \ref{SED}.
The TeV afterglow data are taken from the \cite{2021Sci...372.1081H}.
The optical data are taken from \cite{2019GCN.25569....1C} and \cite{2021A&A...646A..50H}. The spectral hardening is seen around 1 TeV in the first H.E.S.S. observation epoch. Such a spectral feature is also seen in the second H.E.S.S. observation epoch, although the data have large error bars. The radio afterglow data are taken from \cite{2019GCN.25635....1M}.
The X-ray data are taken from Swift/XRT telescope\footnote{Collected from \href{https://www.swift.ac.uk/burst_analyser/}{https://www.swift.ac.uk/}}.
We extract the X-ray spectrum in the temporal coverage of the H.E.S.S. observations.
We fit the spectra with an absorbed single power-law model.
We obtain photon indices of $2.00\pm 0.18$ and $2.2\pm 0.3$  for the first and second H.E.S.S. observation nights, respectively. The derived X-ray $\nu {\rm F}_\nu$ spectra are shown as insets in Figure \ref{SED}.
Since the X-ray photons below 1.5 keV suffered great absorption by the neutral hydrogen in both the GRB host galaxy and our galaxy, we adopt only the X-ray data above 1.5 keV in our SED fits.

\subsection{Leptonic Model Fit to the Multiwavelength Afterglow Data}
We firstly fit the afterglow lightcurves of GRB 190829A in the radio (1.3 GHz, 5 GHz, and 15.5 GHz), optical (u), X-ray (0.3-10 keV), and sub-TeV ($< 0.5$ TeV) gamma-ray bands with our leptonic model only.
Note that the X-ray data at the early epoch of $t<500$ seconds are excluded in our model fitting, since they probably are the residual emission of the prompt emission or X-ray flares.
To account for the extinction effect of the GRB host galaxy, we adopt the Milky Way extinction law with $R_{\rm V}=3.1$ \citep{1999PASP..111...63F}.
A tentative jet break is observed at $\sim 3\times 10^6$ s in the X-ray afterglow lightcurve. We fix the jet break time in our fitting. In addition, the GRB host galaxy extinction parameter is fixed at $E(B-V)=0.64$.
The EBL absorption effect is taken into account for calculating the observed high-energy photons \citep{2008ApJ...686..181F}, $F_{\rm obs}=F_{\rm int}(1-e^{\tau_{\gamma\gamma}})/\tau_{\gamma\gamma}$.

\begin{figure}[ht]
	\centering
	\includegraphics[scale=0.9]{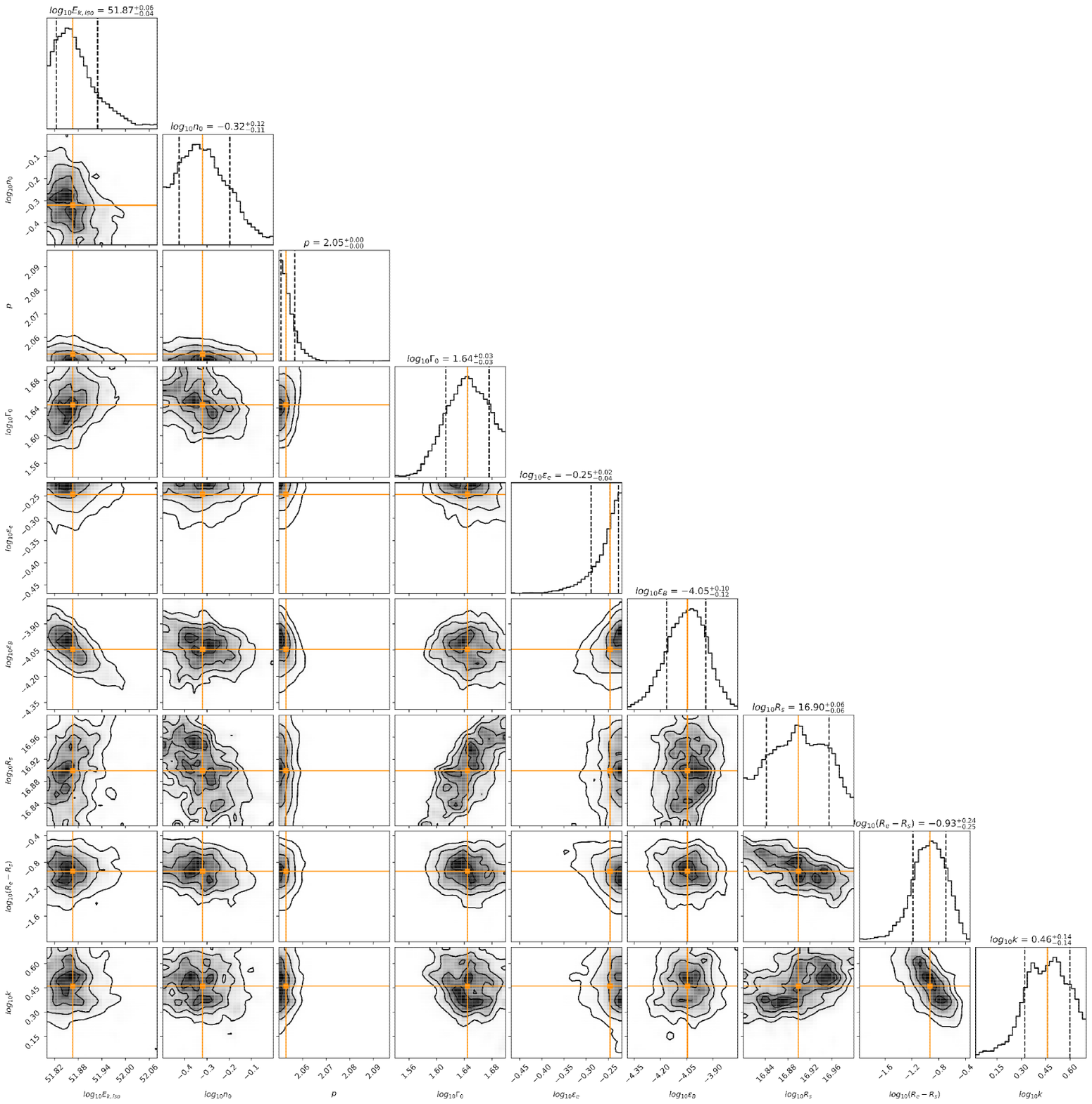}
	\caption{Posterior distribution contours of the leptonic model parameters derived from our MCMC fits for GRB 190829A.}
	\label{MCMC1}
\end{figure}

The {\tt emcee} Python packages \footnote{The package {\tt emcee} is an MCMC algorithm presented in \cite{2013PASP..125..306F}.} are utilized for our joint fitting to the lightcurves and SEDs. The derived probability distributions of the leptonic model parameters are shown in Figure \ref{MCMC1}, which give the initial Lorentz factor as  ${\rm log}_{10}\Gamma_0=1.64\pm 0.03$,
the kinetic energy of the fireball as ${\rm log}_{10}E_{\rm k}/{\rm erg}=51.87^{+0.06}_{-0.04}$,
the energy partition of the magnetic field as ${\rm log}_{10}\epsilon_{\rm B}=-4.05^{+0.10}_{-0.12}$,
the energy partition of electrons as ${\rm log}_{10} \epsilon_{\rm e}=-0.25^{+0.02}_{-0.04}$,
the electron number density as ${\rm log}_{10}n_{\rm 0}/{\rm cm^{-3}}=-0.32^{+0.12}_{-0.11}$,
$p\sim 2.05$, ${\rm log}_{10}R_{\rm s}/{\rm cm}=16.90\pm 0.06$, ${\rm log}_{10}R_{\rm e}/{\rm cm}=17.05$
and ${\rm log}_{10} k = 0.46\pm 0.14$. The inferred jet half-opening angle is ${\rm log}_{10}\theta_{\rm j}=-0.2$ rad by taking $t_{\rm jet}=3\times 10^{6}$ s. The values of the model parameters are generally consistent with those reported in \cite{2021ApJ...917...95Z}.

\begin{figure}[ht]
	\centering
	\includegraphics[scale=0.13]{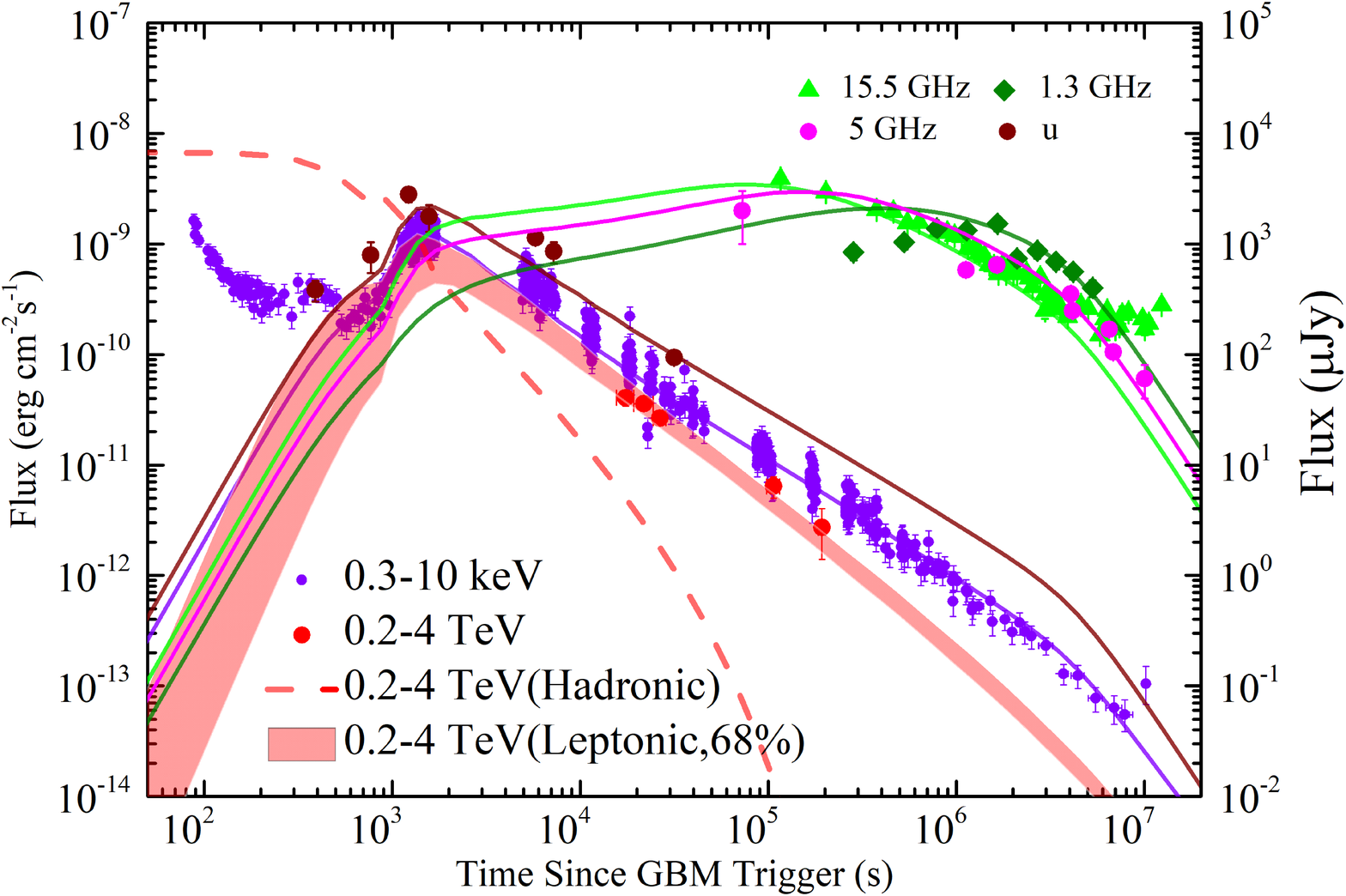}
	\caption{Multiwavelength afterglow lightcurves of GRB 190829A (points) and the best fits of our lepto-hadronic model (lines). The TeV gamma-ray flux is corrected by the EBL absorption.}
    \label{LC}
\end{figure}

The fitting curves are also reported in Figures \ref{LC} and \ref{SED}.
The afterglow lightcurves in the gamma-ray (0.2-4 TeV), X-ray (0.3-10 keV), optical, and high-frequency radio bands can be well fit with our models.
The observed radio fluxes at 1.5 GHz and 5 GHz at $t<1.3\times 10^{6}$ s are lower than our model prediction.
The uncertainties of the 0.2-4 TeV lightcurve and the SEDs are also illustrated as bands in Figures \ref{LC} and \ref{SED}.
These uncertainties are plotted by randomly selecting 800 model parameter sets that yield convergence results in our MCMCMarkov Chain Monte
Carlo (MCMC) fit.
The values of the model parameters are in the 68\% ranges ($1\sigma$) of their posterior probability distributions.
It can be seen that leptonic model can fit the observed lightcurves and SEDs if we consider the $1\sigma$ uncertainties of the model parameters.

\subsection{Constraining the UHE Proton Population with the Lepto-hadronic Model}

\begin{figure}[ht]
	\centering
	\includegraphics[scale=0.1]{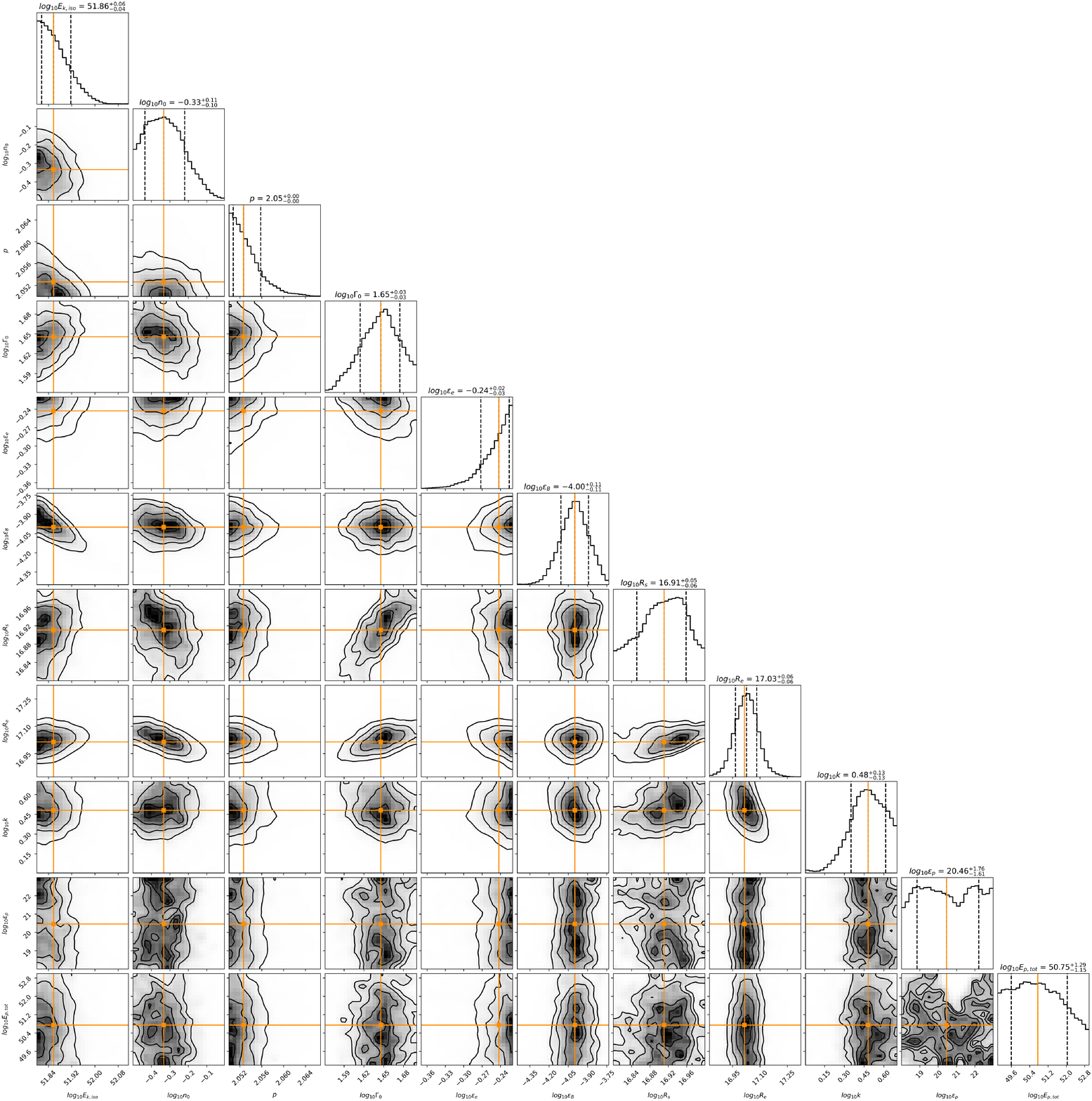}
	\caption{Posterior distribution contours of the lepto-hadronic model parameters derived from our MCMC fits for GRB 190829A.}
	\label{MCMC2}
\end{figure}

The hard TeV afterglow spectrum of GRB 190829A motivates us to fit its multiwavelength afterglow lightcurves and SEDs with our lepto-hadronic model. We attribute the VHE gamma-ray afterglows observed with H.E.S.S. to the SSC process of the accelerated electrons accelerated in situ by external shocks and the synchrotron radiations of the UHE protons pre-accelerated by the internal shocks. We find that the lepto-hadronic model does not improve the fits. The derived probability distributions of the model parameters from our fit are shown in Figure \ref{MCMC2}, which give ${\rm log}_{10}\Gamma_0=1.65\pm 0.03$,
${\rm log}_{10}E_{\rm k}/{\rm erg}=51.86^{+0.06}_{-0.04}$,
${\rm log}_{10}\epsilon_{\rm B}=-4.00^{+0.11}_{-0.11}$,
${\rm log}_{10} \epsilon_{\rm e}=-0.24^{+0.02}_{-0.03}$,
${\rm log}_{10}n_{\rm 0}/{\rm cm^{-3}}=-0.33^{+0.11}_{-0.10}$, $p\sim 2.05$,
${\rm log}_{10}R_{\rm s}/{\rm cm}=16.91^{+0.05}_{-0.06}$, ${\rm log}_{10}R_{\rm e}/{\rm cm}=17.03^{+0.06}_{-0.06}$
and ${\rm log}_{10} k = 0.48^{+0.13}_{-0.13}$.
They are consistent with that derived by considering the leptonic model only at the $1\sigma$ confidence level, as shown in Figure \ref{MCMC1}. Therefore, one cannot statistically claim the detection of an extra emission component over the leptonic model.

Indeed, the parameters of the hadronic component are poorly constrained in our MCMC fit, i.e. $\log_{10}\varepsilon_{\rm p}^{\prime}/{\rm eV}\sim 20.46$ and the total energy of the proton population $\log_{10}E_{\rm p, total}/{\rm erg}\leq 50.75$. From their posterior probability distributions, we have $\log_{10}\varepsilon_{\rm p}^{\prime}/{\rm eV}=20.46^{+1.76}_{-1.61}$ and  $\log_{10}E_{\rm p, total}/{\rm erg}=50.75^{+1.29}_{-1.15}$. We generate the global profile of the UHE proton synchrotron emission by randomly selecting 800 model parameter sets. The selected parameters are in the 68\% ranges ($1\sigma$) of their posterior probability distributions and yield convergence results in our MCMC fit. The upper boundary of the profile, which could be regarded as the upper limit of the UHE proton emission derived from the MCMC fit, is also shown in Figure \ref{SED}. Note that the upper limit is not a smoothed curve, since the selected $\{\varepsilon_{\rm p}^{\prime}, E_{\rm p, total}\}$ sets are in their $1\sigma$ posterior probability distributions. One can find that the TeV spectrum observed in the first H.E.S.S. observation night could potentially be hardened by the UHE proton emission, although the TeV data can still be explained by the leptonic model.

Adopting ${\rm log}_{10}\varepsilon_{\rm p}^{\prime}/{\rm eV}=20.46$ and $\log_{10}E_{\rm p, total}/{\rm erg}=50.75$, we calculate the 0.2-4 TeV gamma-ray afterglow lightcurves in the observer frame for the proton synchrotron emission. They are also shown in Figure \ref{LC}. The observed TeV gamma-ray afterglows with H.E.S.S. at late epochs ($t>10^3$ s) are dominated by the electron SSC process. The proton synchrotron emission dominates the observed TeV flux at the early epoch, but it decays rapidly with a slope of $\sim -2.12$ post the fireball deceleration time. This is due to the lack of fresh, continuously injected UHE protons as the fireball expands. It is interesting that the decay of the TeV gamma-ray flux from the UHE protons after the deceleration timescale is similar to that of the reverse shock emission. However, it is almost a constant before the deceleration timescale of the fireball. This is different from the reverse shock emission. Very early observation is critical for identifying the hadronic component.

\section{Discussion}
\subsection{The Saturation Energies of Protons and Electrons accelerated in the Prompt Emission Phase}
As proposed by \cite{1995PhRvL..75..386W}, protons can be accelerated to UHE in the internal shocks and dissipate their kinetic energy at a radius of $\geq 10^{13}$ cm. The maximum proton energy accelerated in the shocks is estimated through $t^{'}_{\rm acc} = t^{'}_{\rm syn}$, where $t^{'}_{\rm acc}$ and $t^{'}_{\rm syn}$ are the acceleration and the synchrotron radiation timescales of the protons, respectively.
Thus, the maximum proton energy ($\varepsilon^{'}_{\rm p,max}$) in the comoving frame can be derived as (e.g. \citealp{2011MNRAS.418.1382L})
\begin{equation}
\varepsilon^{'}_{\rm p,max}\simeq 9.7 \times 10^{19}~\eta^{-1/2}\epsilon^{1/4}_{\rm e,-1}\epsilon^{-1/4}_{\rm B,-1}L^{-1/4}_{\gamma,\rm iso,47}\Gamma^{5/2}_{0,1}{\delta t_{2}^{1/2}} \beta_{\rm sh} {\rm eV},
\end{equation}
where $\delta t$ is the minimum variability timescale of the prompt gamma rays, $\eta (\geq 1)$ is the so-called gyro-factor \citep{1999APh....11..347H,2000NewA....5..377A} and $\beta_{\rm sh}$ is the velocity of the internal shocks in units of $c$. As discussed in \cite{1994MNRAS.269L..41M} and \cite{1995Ap&SS.231..441M}, $\beta_{\rm sh}$ is about 0.1-0.3 in the  mildly relativistic internal shocks.
Taking $\delta t=0.214$ s (\citealp{2021ApJ...918...12F}), $\eta=1$, $\beta_{\rm sh} \approx 0.3$, $L_{\gamma,\rm iso}=2.6\times10^{49}$ erg s$^{-1}$, and the values of $\epsilon_{\rm e}$, $\epsilon_{\rm B}$, and $\Gamma_0$ derived from our fit to the multiple afterglow lightcurves, we have  $\varepsilon^{'}_{\rm p,\max}=1.23\times10^{20}{\rm eV}$, which is of the same order as the upper limit of $\varepsilon_{\rm p}^{'}$ derived from our analysis above.

Note that the most restrictive requirement of the particle acceleration in a source is that the Larmor radius ($R_{\rm L}$) of  particles should be smaller than the internal shock radius $R_{\rm sh}$, which is estimated as $R_{\rm sh}=2\Gamma_{0}^{2}c\delta t=2.56\times10^{13}{\rm cm}$.
$R_{\rm L}$ is calculated as $R_{\rm L}=\frac{\varepsilon^{'}_{\rm p,\max}}{\Gamma_{0}eB_{\rm int}^{\prime}}$, where $B_{\rm int}^{\prime}=(8\pi U_{\rm B})^{1/2}$ and $U_{\rm B}$ is the energy density of the magnetic field.
We estimate $U_{\rm B}$ as $U_{\rm B}=\frac{\epsilon_{\rm B}}{\epsilon_{\rm e}}U_{\gamma}$, where $U_{\gamma}=L_{\gamma,\rm iso}/4\pi R_{\rm sh}^{2}\Gamma_0^{2}c$ is the comoving gamma-ray energy density.
We obtain $B_{\rm int}^{\prime}\sim 4.80\times 10^2$ G and $R_{\rm L}=4.48\times10^{13}{\rm cm}$ for protons with $\log_{10} \varepsilon^{'}_{\rm p,\max}/{\rm eV}=20.46$.
It is found that $R_{\rm L}$ is at the same order as $R_{\rm sh}$.
These UHE protons may have the possibility to being accelerated in the internal shock region.

Similarly, we also estimate the saturation energy of the electrons ($\gamma_{\rm e,s}m_{\rm e} c^2$) in the internal shocks with the condition of $t^{'}_{\rm acc}=t^{'}_{\rm syn,e}$, where $t^{'}_{\rm acc,e}=\eta\frac{\gamma_{\rm e} m_{\rm e}c}{eB_{\rm int}^{\prime}\Gamma_{0}\beta_{\rm sh}^{2}}$ and $t^{'}_{\rm syn,e}=\frac{6\pi m_{\rm e}c}{\gamma_{\rm e}\sigma_{\rm T}B_{\rm int}^{\prime,2}}$.
We have $\gamma_{\rm e,s}=1.17\times10^{7}\eta^{-1/2}B_{\rm int, 2}^{\prime,-1/2}\beta_{\rm sh}\Gamma^{1/2}_{0}\simeq 1.60\times 10^6$ for $\beta_{\rm sh}=0.3$, $\eta=1$, and $B_{\rm int}^{\prime}=4.80\times 10^2$ G.
The synchrotron radiations of these electrons in the observer frame peak at $\varepsilon_{\gamma}=2\Gamma_0 h\nu(\gamma_{\rm e,s})/(1+z)=2\Gamma_0\gamma_{\rm e,s}^{2}\frac{heB_{\rm int}^{\prime}}{2\pi m_{\rm e}c}/(1+z)=2.31\times10^{9}B_{2}\Gamma_{0, 1}\gamma_{\rm e,s,7}^{2}(1+z)^{-1}\simeq 1.18$ GeV.
Since there have been no prompt GeV gamma-ray observations, we cannot place any constraints on this electron population in the jet of GRB 190829A.

\subsection{Hadronic Processes of VHE Protons in the Afterglow Phase}
In our model, the pre-accelerated UHE protons in the prompt gamma-ray phase are assumed to be injected into the afterglow fireball.
Whether or not the UHE protons can be confined in the afterglow model is a great issue.
Note that $R_L$ rapidly increases as the magnetic field strength and the Lorentz factor $\Gamma$ of the fireball quickly decay during the expansion of the fireball, since $R_{\rm L}=\frac{\varepsilon^{'}_{\rm p,max}}{\Gamma eB_{\rm ext}^{\prime}}$.
Based on our analysis results, we have $B_{\rm ext}^{\prime}=1.73\times10^{-2} {\rm ~G}$, $\Gamma=7.01$, and $R=2.43\times10^{17}$ cm at $t \sim 2\times 10^{4}{\rm s}$ (the first night of the H.E.S.S. observation).
The corresponding $R_{\rm L}$ of the UHE protons is $5.55\times 10^{19}{\,\rm cm}$, which is about two orders of magnitude larger than $R$.
Therefore, the UHE protons cannot be confined and continuously accelerated in the external shock fireball.
It is possible that the UHE protons could be cooled by synchrotron radiations during their escape from the external fireball.
The maximum escape timescale of the UHE protons in the afterglow fireball is estimated as $t_{\rm esc}\simeq R(t)/\Gamma^2 c$.
Note that $ R(t)=\kappa \Gamma^{2} c t$, where the numerical factor $\kappa$ varies between $3\sim 7$ depending on the details of the hydrodynamic evolution and the spectrum \citep{1997ApJ...482..942P,1997ApJ...489L..37S,1997ApJ...491L..19W,1998ApJ...497L..17S}.
Thus, we have $t_{\rm esc}\simeq \kappa t>t$.
This indicates that a fraction of the UHE protons can lose their energy via the synchrotron radiations before their escape out of the external fireball, even at the very early stage of the afterglow fireball.
Our numerical calculation gives $t_{\rm esc}=1.63\times10^{5}$ s at $t=2\times 10^4$ s.
\cite{2022ApJ...929...70S} proposed $p-\gamma$ process model to explain the TeV emission of GRB 190829A by considering the SSC photons as the seed photons.
To yield a gamma-ray luminosity of $\sim 10^{44}$ erg s$^{-1}$ (the VHE gamma-ray luminosity in the first night H.E.S.S. observation of GRB 190829A), they found that the density of the seed photons is $\sim 10^8$ cm$^{-3}$, which is six orders of magnitude higher than the photon field of the SSC photons based on our leptonic model.
Therefore, if the observed VHE gamma-ray flux is attributed to the $p-\gamma$ process, extremely dense seed photons are required.

Our above analysis considers only the synchrotron radiations of the UHE protons pre-accelerated in the prompt gamma-ray phase.
Their emissions from the neutral pion decay in the $p-p$ and $p-\gamma$ interactions \citep{2006PhRvD..74c4018K,2008PhRvD..78c4013K} are ignored, since the cooling efficiencies of the the $p-p$ and $p-\gamma$ interactions of the UHE protons are extremely low and the emitted photons are out of the H.E.S.S. energy range.
We should note that the protons in the afterglow fireball could be accelerated in situ by the external shocks when the external shocks propagate into the surrounding medium.
We consider the emission from the $p-\gamma$ process of these protons.
The seed photons are from the electron synchrotron radiations derived from our leptonic model.
Assuming that the protons are co-accelerated with the electrons, we take the power-law energy distribution of the protons in the range from $\varepsilon^{\rm a,'}_{\rm p,min}$ to $\varepsilon^{\rm a,'}_{\rm p,max}$ to be the same as the electrons, i.e. $p_{\rm p}=2.05$, where $\varepsilon^{\rm a,'}_{\rm p,min}$ is the proton rest energy and $\varepsilon^{\rm a,'}_{\rm p,max}$ is estimated as $\varepsilon^{\rm a,'}_{\rm p,max}=\eta^{-1}ceB_{\rm ext}^{\prime}t^{'}_{\rm acc}=7.35\times 10^{14} {~\rm eV}~ \eta^{-1} B_{\rm ext, -1}^{\prime} t^{'}_{\rm acc,3}=2.54\times 10^{15}{~\rm eV}$.
The number density of the protons is also the same as the electrons, i.e. ${\rm log}_{10}n_{\rm p}/{\rm cm^{-3}}=-0.33$.
The total energy carried by the protons is $E_{\rm p, total}=3.43\times10^{50}{\rm erg}$, being comparable to the $E_{\rm k}$ value derived from our leptonic model fit.
We calculate the emission from the $p-\gamma$ interaction at $t=2\times 10^4$ s and $t=10^5$ s by adopting the same model as presented in \cite{2008PhRvD..78c4013K}.
It is found that the derived gamma-ray flux is $1.01\times10^{-18}$ erg cm$^{-2}$ s$^{-1}$ in the 0.2-4 TeV band, which is about 7 orders of magnitude lower than that from the synchrotron radiations of the pre-accelerated UHE protons.
For the $p-p$ process, we take the high-energy protons to be the same as in the $p-\gamma$ process.
In an electrically neutral environment, the density of low-energy protons is equal to the electron number density, ${\rm log}_{10}n_{\rm e}/{\rm cm^{-3}}=-0.33$.
Thus, the radiation efficiency of $p-p$ process is lower than the $p-\gamma$ process.
Therefore, both the $p-\gamma$ and $p-p$ processes of the VHE protons accelerated in situ by the external shocks cannot explain the observed H.E.S.S. data.

VHE gamma-ray photons are messengers of UHE protons.
BATSE/EGRET firstly detected an 18 GeV photon from GRB 940217 arrived after 75 minutes of the burst \citep{1994Natur.372..652H}, and Fermi/LAT confirmed that a considerable fraction of GRBs are indeed GeV emitters \citep{2011ApJ...730..141Z}.
The search for TeV gamma rays of GRBs has been pioneered by some groups \citep{1998A&A...337...43P, 2017ApJ...842...31B}, and the detection of TeV gamma rays from GRB 190829A \citep{2021Sci...372.1081H} offers plausible evidence of $\sim 10^{20}$ eV protons accelerated in the GRB jet. 

Neutrinos radiated via the hadronic processes are other messengers of UHE proton cooling.
A dedicated search for neutrinos in space and time coincident with GRB 190829A has been performed using ANTARES data, but no positive results has been found \citep{2021JCAP...03..092A}.
We estimate the peak flux of the neutrinos from the $p-\gamma$ process as $\nu F_{\nu_{\rm \mu, peak}} \thickapprox 10^{-16}$ erg  cm$^{-2}$ s$^{-1}$ at ${\varepsilon_{\mu}}\thickapprox10^{14}{\rm eV}$, which is about five orders lower than the threshold of the current neutrino detectors \citep{2020PhRvL.124e1103A}.

\section{Summary}
\label{sec:Con}
We have presented theoretical fits to the multiwavelength afterglow data of GRB 190829A for exploring possible signatures of synchrotron radiation from the pre-accelerated UHE protons in its TeV gamma-ray data. We summarize our results as follows.

First, considering the leptonic model only, we show that the leptonic model in $1\sigma$ confidence level of its model parameters can represent the observed multiwavelength lightcurves and SEDs of GRB 190829A within the error bars of the data. The derived model parameters are  ${\rm log}_{10}\Gamma_0=1.64\pm 0.03$, ${\rm log}_{10}E_{\rm k}/{\rm erg}=51.87^{+0.06}_{-0.04}$,
${\rm log}_{10}\epsilon_{\rm B}=-4.05^{+0.10}_{-0.12}$,
${\rm log}_{10} \epsilon_{\rm e}=-0.25^{+0.02}_{-0.04}$,
${\rm log}_{10}n_{\rm 0}/{\rm cm^{-3}}=-0.32^{+0.12}_{-0.11}$,
$p\sim 2.05$, ${\rm log}_{10}R_{\rm s}/{\rm cm}=16.90\pm 0.06$, ${\rm log}_{10}R_{\rm e}/{\rm cm}=17.05$, ${\rm log}_{10} k = 0.46\pm 0.14$, and ${\rm log}_{10}\theta_{\rm j}=-0.2$ rad (fixed).

Second, attributing the TeV gamma-ray afterglows to the emission of both the electron self-Compton scattering process and the synchrotron radiation of the UHE proton population in the afterglow fireball, our lepto-hadronic model does not improve fits to the data over the leptonic model, indicating that the UHE proton component cannot be statistically claimed. Our MCMC fit yields upper limits of the UHE proton population as $\log_{10} \varepsilon_{\rm p}^{\prime}/{\rm eV}\sim 20.46$ and $\log_{10}E_{\rm p, total}/{\rm erg}\leq 50.75$.

We have also presented a brief discussion on the saturation energies of protons and electrons accelerated in the internal shocks of GRB 190829A. we show that the UHE protons are potentially accelerated in the prompt gamma-ray phase. Since the TeV gamma-ray flux of the proton synchrotron radiations predicted by our model dominates the early TeV flux at $t<10^3$ s but decays rapidly with a slope of $\sim -2.12$ post the deceleration of the fireball, early TeV gamma-ray observations are critical for revealing the UHE proton population. The saturation energy of the electrons could be up to $1.6\times 10^6$. The synchrotron emission of these electrons peaks at 1.18 GeV, but we cannot place any constraints on this electron population, since we have no GeV observations of the prompt phase of GRB 190829A. The $p-p$ and $p-\gamma$ hadronic processes of VHE protons accelerated in the afterglow phase are also discussed.

\acknowledgments
We appreciate the very thoughtful and valuable comments from the referee. We are thankful for the helpful discussions with Xiangyu Wang, Jin Zhang, Da-Bin Lin, Bo-hua Li, and Kuan Liu.
This work is supported by the National Natural Science Foundation of China (grant Nos.12133003 and 12203015) and Innovation Project of Guangxi Graduate Education (YCBZ2021025).

\clearpage

\begin{thebibliography}{}
	\expandafter\ifx\csname natexlab\endcsname\relax\def\natexlab#1{#1}\fi
	\providecommand{\url}[1]{\href{#1}{#1}}
	\providecommand{\dodoi}[1]{doi:~\href{http://doi.org/#1}{\nolinkurl{#1}}}
	\providecommand{\doeprint}[1]{\href{http://ascl.net/#1}{\nolinkurl{http://ascl.net/#1}}}
	\providecommand{\doarXiv}[1]{\href{https://arxiv.org/abs/#1}{\nolinkurl{https://arxiv.org/abs/#1}}}

\bibitem[Aartsen et al.(2020)]{2020PhRvL.124e1103A} Aartsen, M.~G., Ackermann, M., Adams, J., et al.\ 2020, \prl, 124, 051103. \dodoi{10.1103/PhysRevLett.124.051103}
\bibitem[Ackermann et al.(2014)]{2014Sci...343...42A} Ackermann, M., Ajello, M., Asano, K., et al.\ 2014, Science, 343, 42. \dodoi{10.1126/science.1242353}
\bibitem[Aharonian(2000)]{2000NewA....5..377A} Aharonian, F.~A.\ 2000, \na, 5, 377. \dodoi{10.1016/S1384-1076(00)00039-7}
\bibitem[Alves et al.(2018)]{2018PhRvL.121x5101A} Alves, E.~P., Zrake, J., \& Fiuza, F.\ 2018, \prl, 121, 245101. \dodoi{10.1103/PhysRevLett.121.245101}
\bibitem[ANTARES Collaboration et al.(2021)]{2021JCAP...03..092A} ANTARES Collaboration, Albert, A., Andr{\'e}, M., et al.\ 2021, \jcap, 2021, 092. \dodoi{10.1088/1475-7516/2021/03/092}
\bibitem[Bartoli et al.(2017)]{2017ApJ...842...31B} Bartoli, B., Bernardini, P., Bi, X.~J., et al.\ 2017, \apj, 842, 31. \dodoi{10.3847/1538-4357/aa74bc}
\bibitem[Blandford \& Eichler(1987)]{1987PhR...154....1B} Blandford, R. \& Eichler, D.\ 1987, \physrep, 154, 1. \dodoi{10.1016/0370-1573(87)90134-7}
\bibitem[Chand et al.(2020)]{2020ApJ...898...42C} Chand, V., Banerjee, A., Gupta, R., et al.\ 2020, \apj, 898, 42. \dodoi{10.3847/1538-4357/ab9606}
\bibitem[Chen et al.(2019)]{2019GCN.25569....1C} Chen, T.-W., Bolmer, J., Nicuesa Guelbenzu, A., et al.\ 2019, GRB Coordinates Network, Circular Service, No. 25569, 1
\bibitem[Crusius \& Schlickeiser(1986)]{1986A&A...164L..16C} Crusius, A. \& Schlickeiser, R.\ 1986, \aap, 164, L16
\bibitem[Derishev \& Piran(2019)]{2019ApJ...880L..27D} Derishev, E. \& Piran, T.\ 2019, \apjl, 880, L27. \dodoi{10.3847/2041-8213/ab2d8a}
\bibitem[Drury et al.(1999)]{1999A&A...347..370D} Drury, L.~O., Duffy, P., Eichler, D., et al.\ 1999, \aap, 347, 370
\bibitem[Duan \& Wang(2019)]{2019ApJ...884...61D} Duan, M.-Y. \& Wang, X.-G.\ 2019, \apj, 884, 61. \dodoi{10.3847/1538-4357/ab3c6e}
\bibitem[Fan et al.(2008)]{2008MNRAS.384.1483F} Fan, Y.-Z., Piran, T., Narayan, R., et al.\ 2008, \mnras, 384, 1483. \dodoi{10.1111/j.1365-2966.2007.12765.x}
\bibitem[Finke et al.(2008)]{2008ApJ...686..181F} Finke, J.~D., Dermer, C.~D., \& B{\"o}ttcher, M.\ 2008, \apj, 686, 181. \dodoi{10.1086/590900}
\bibitem[Fitzpatrick(1999)]{1999PASP..111...63F} Fitzpatrick, E.~L.\ 1999, \pasp, 111, 63. \dodoi{10.1086/316293}
\bibitem[Foreman-Mackey et al.(2013)]{2013PASP..125..306F} Foreman-Mackey, D., Hogg, D.~W., Lang, D., et al.\ 2013, \pasp, 125, 306. \dodoi{10.1086/670067}
\bibitem[Fraija et al.(2019)]{2019ApJ...885...29F} Fraija, N., Dichiara, S., Pedreira, A.~C.~C. do E.~S., et al.\ 2019, \apj, 885, 29. \dodoi{10.3847/1538-4357/ab3e4b}
\bibitem[Fraija et al.(2021)]{2021ApJ...918...12F} Fraija, N., Veres, P., Beniamini, P., et al.\ 2021, \apj, 918, 12. \dodoi{10.3847/1538-4357/ac0aed}
\bibitem[Gould \& Schr{\'e}der(1966)]{1966PhRvL..16..252G} Gould, R.~J. \& Schr{\'e}der, G.\ 1966, \prl, 16, 252. \dodoi{10.1103/PhysRevLett.16.252}
\bibitem[H.E.S.S. Collaboration et al.(2021)]{2021Sci...372.1081H} H.~E.~S.~S. Collaboration, Abdalla, H., Aharonian, F., et al.\ 2021, Science, 372, 1081. \dodoi{10.1126/science.abe8560}
\bibitem[Henri et al.(1999)]{1999APh....11..347H} Henri, G., Pelletier, G., Petrucci, P.~O., et al.\ 1999, Astroparticle Physics, 11, 347. \dodoi{10.1016/S0927-6505(98)00071-1}
\bibitem[Hu et al.(2021)]{2021A&A...646A..50H} Hu, Y.-D., Castro-Tirado, A.~J., Kumar, A., et al.\ 2021, \aap, 646, A50. \dodoi{10.1051/0004-6361/202039349}
\bibitem[Huang et al.(2020)]{2020ApJ...903L..26H} Huang, X.-L., Liang, E.-W., Liu, R.-Y., et al.\ 2020, \apjl, 903, L26. \dodoi{10.3847/2041-8213/abc330}
\bibitem[Huang et al.(2022)]{2022ApJ...925..182H} Huang, Z.-Q., Kirk, J.~G., Giacinti, G., et al.\ 2022, \apj, 925, 182. \dodoi{10.3847/1538-4357/ac3f38}
\bibitem[Hurley et al.(1994)]{1994Natur.372..652H} Hurley, K., Dingus, B.~L., Mukherjee, R., et al.\ 1994, \nat, 372, 652. \dodoi{10.1038/372652a0}
\bibitem[Kelner \& Aharonian(2008)]{2008PhRvD..78c4013K} Kelner, S.~R. \& Aharonian, F.~A.\ 2008, \prd, 78, 034013. \dodoi{10.1103/PhysRevD.78.034013}
\bibitem[Kelner et al.(2006)]{2006PhRvD..74c4018K} Kelner, S.~R., Aharonian, F.~A., \& Bugayov, V.~V.\ 2006, \prd, 74, 034018. \dodoi{10.1103/PhysRevD.74.034018}
\bibitem[Kumar \& Zhang(2015)]{2015PhR...561....1K} Kumar, P. \& Zhang, B.\ 2015, \physrep, 561, 1. \dodoi{10.1016/j.physrep.2014.09.008}
\bibitem[Liang et al.(2006)]{2006ApJ...646..351L} Liang, E.~W., Zhang, B., O'Brien, P.~T., et al.\ 2006, \apj, 646, 351. \dodoi{10.1086/504684}
\bibitem[Liu et al.(2011)]{2011MNRAS.418.1382L} Liu, R.-Y., Wang, X.-Y., \& Dai, Z.-G.\ 2011, \mnras, 418, 1382. \dodoi{10.1111/j.1365-2966.2011.19590.x}
\bibitem[Liu et al.(2013)]{2013ApJ...773L..20L} Liu, R.-Y., Wang, X.-Y., \& Wu, X.-F.\ 2013, \apjl, 773, L20. \dodoi{10.1088/2041-8205/773/2/L20}
\bibitem[MAGIC Collaboration et al.(2019a)]{2019Natur.575..455M} MAGIC Collaboration, Acciari, V.~A., Ansoldi, S., et al.\ 2019a, \nat, 575, 455. \dodoi{10.1038/s41586-019-1750-x}
\bibitem[MAGIC Collaboration et al.(2020)]{2020A&A...637A..86M} MAGIC Collaboration, Acciari, V.~A., Ansoldi, S., et al.\ 2020, \aap, 637, A86. \dodoi{10.1051/0004-6361/201834603}
\bibitem[MAGIC Collaboration et al.(2019b)]{2019Natur.575..459M} MAGIC Collaboration, Acciari, V.~A., Ansoldi, S., et al.\ 2019b, \nat, 575, 459. \dodoi{10.1038/s41586-019-1754-6}
\bibitem[Medina-Torrej{\'o}n et al.(2021)]{2021ApJ...908..193M} Medina-Torrej{\'o}n, T.~E., de Gouveia Dal Pino, E.~M., Kadowaki, L.~H.~S., et al.\ 2021, \apj, 908, 193. \dodoi{10.3847/1538-4357/abd6c2}
\bibitem[Melrose \& Crouch(1997)]{1997PASA...14..251M} Melrose, D. \& Crouch, A.\ 1997, \pasa, 14, 251. \dodoi{10.1071/AS97251}
\bibitem[Meszaros \& Rees(1994)]{1994MNRAS.269L..41M} Meszaros, P. \& Rees, M.~J.\ 1994, \mnras, 269, L41. \dodoi{10.1093/mnras/269.1.L41}
\bibitem[Meszaros et al.(1994)]{1994ApJ...432..181M} Meszaros, P., Rees, M.~J., \& Papathanassiou, H.\ 1994, \apj, 432, 181. \dodoi{10.1086/174559}
\bibitem[Meszaros \& Rees(1993)]{1993ApJ...418L..59M} Meszaros, P. \& Rees, M.~J.\ 1993, \apjl, 418, L59. \dodoi{10.1086/187116}
\bibitem[Milgrom \& Usov(1995)]{1995ApJ...449L..37M} Milgrom, M. \& Usov, V.\ 1995, \apjl, 449, L37. \dodoi{10.1086/309633}
\bibitem[Mochkovitch et al.(1995)]{1995Ap&SS.231..441M} Mochkovitch, R., Maitia, V., \& Marques, R.\ 1995, \apss, 231, 441. \dodoi{10.1007/BF00658666}
\bibitem[Monageng et al.(2019)]{2019GCN.25635....1M} Monageng, I., van der Horst, A.~J., Woudt, P.~A., et al.\ 2019, GRB Coordinates Network, Circular Service, No. 25635, 1
\bibitem[Murase \& Nagataki(2006)]{2006PhRvD..73f3002M} Murase, K. \& Nagataki, S.\ 2006, \prd, 73, 063002. \dodoi{10.1103/PhysRevD.73.063002}
\bibitem[M{\'e}sz{\'a}ros \& Rees(1997)]{1997ApJ...482L..29M} M{\'e}sz{\'a}ros, P. \& Rees, M.~J.\ 1997, \apjl, 482, L29. \dodoi{10.1086/310692}
\bibitem[Padilla et al.(1998)]{1998A&A...337...43P} Padilla, L., Funk, B., Krawczynski, H., et al.\ 1998, \aap, 337, 43
\bibitem[Panaitescu et al.(1997)]{1997ApJ...482..942P} Panaitescu, A., Wen, L., Laguna, P., et al.\ 1997, \apj, 482, 942. \dodoi{10.1086/304185}
\bibitem[Piran(2004)]{2004RvMP...76.1143P} Piran, T.\ 2004, Reviews of Modern Physics, 76, 1143. \dodoi{10.1103/RevModPhys.76.1143}
\bibitem[Protheroe \& Stanev(1999)]{1999APh....10..185P} Protheroe, R.~J. \& Stanev, T.\ 1999, Astroparticle Physics, 10, 185. \dodoi{10.1016/S0927-6505(98)00055-3}
\bibitem[Rachen \& Biermann(1993)]{1993A&A...272..161R} Rachen, J.~P. \& Biermann, P.~L.\ 1993, \aap, 272, 161
\bibitem[Ren et al.(2020)]{2020ApJ...901L..26R} Ren, J., Lin, D.-B., Zhang, L.-L., et al.\ 2020, \apjl, 901, L26. \dodoi{10.3847/2041-8213/abb672}
\bibitem[Ren et al.(2022)]{2022arXiv221010673R} Ren, J., Wang, Y., \& Zhang, L.-L.\ 2022, \doarXiv{2210.10673}
\bibitem[Rhodes et al.(2020)]{2020MNRAS.496.3326R} Rhodes, L., van der Horst, A.~J., Fender, R., et al.\ 2020, \mnras, 496, 3326. \dodoi{10.1093/mnras/staa1715}
\bibitem[Rudolph et al.(2022)]{2022MNRAS.511.5823R} Rudolph, A., Bo{\v{s}}njak, {\v{Z}}., Palladino, A., et al.\ 2022, \mnras, 511, 5823. \dodoi{10.1093/mnras/stac433}
\bibitem[Sahu et al.(2022)]{2022ApJ...929...70S} Sahu, S., Valadez Polanco, I.~A., \& Rajpoot, S.\ 2022, \apj, 929, 70. \dodoi{10.3847/1538-4357/ac5cc6}
\bibitem[Salafia et al.(2022)]{2022ApJ...931L..19S} Salafia, O.~S., Ravasio, M.~E., Yang, J., et al.\ 2022, \apjl, 931, L19. \dodoi{10.3847/2041-8213/ac6c28}
\bibitem[Sari(1997)]{1997ApJ...489L..37S} Sari, R.\ 1997, \apjl, 489, L37. \dodoi{10.1086/310957}
\bibitem[Sari et al.(1998)]{1998ApJ...497L..17S} Sari, R., Piran, T., \& Narayan, R.\ 1998, \apjl, 497, L17. \dodoi{10.1086/311269}
\bibitem[Sari \& Esin(2001)]{2001ApJ...548..787S} Sari, R. \& Esin, A.~A.\ 2001, \apj, 548, 787. \dodoi{10.1086/319003}
\bibitem[Stawarz \& Petrosian(2008)]{2008ApJ...681.1725S} Stawarz, {\L}. \& Petrosian, V.\ 2008, \apj, 681, 1725. \dodoi{10.1086/588813}
\bibitem[Totani(1998b)]{1998ApJ...509L..81T} Totani, T.\ 1998, \apjl, 509, L81. \dodoi{10.1086/311772}
\bibitem[Totani(1998a)]{1998ApJ...502L..13T} Totani, T.\ 1998, \apjl, 502, L13. \dodoi{10.1086/311489}
\bibitem[Tramacere et al.(2011)]{2011ApJ...739...66T} Tramacere, A., Massaro, E., \& Taylor, A.~M.\ 2011, \apj, 739, 66. \dodoi{10.1088/0004-637X/739/2/66}
\bibitem[Vietri(1995)]{1995ApJ...453..883V} Vietri, M.\ 1995, \apj, 453, 883. \dodoi{10.1086/176448}
\bibitem[Virtanen \& Vainio(2005)]{2005ApJ...621..313V} Virtanen, J.~J.~P. \& Vainio, R.\ 2005, \apj, 621, 313. \dodoi{10.1086/427324}
\bibitem[Wang et al.(2019)]{2019ApJ...884..117W} Wang, X.-Y., Liu, R.-Y., Zhang, H.-M., et al.\ 2019, \apj, 884, 117. \dodoi{10.3847/1538-4357/ab426c}
\bibitem[Waxman(1997)]{1997ApJ...491L..19W} Waxman, E.\ 1997, \apjl, 491, L19. \dodoi{10.1086/311057}
\bibitem[Waxman(1995)]{1995PhRvL..75..386W} Waxman, E.\ 1995, \prl, 75, 386. \dodoi{10.1103/PhysRevLett.75.386}
\bibitem[Zhang et al.(2021a)]{2021ApJ...920...55Z} Zhang, B.~T., Murase, K., Veres, P., et al.\ 2021a, \apj, 920, 55. \dodoi{10.3847/1538-4357/ac0cfc}
\bibitem[Zhang et al.(2011)]{2011ApJ...730..141Z} Zhang, B.-B., Zhang, B., Liang, E.-W., et al.\ 2011, \apj, 730, 141. \dodoi{10.1088/0004-637X/730/2/141}
\bibitem[Zhang(2019)]{2019Natur.575..448Z} Zhang, B.\ 2019, \nat, 575, 448. \dodoi{10.1038/d41586-019-03503-6}
\bibitem[Zhang et al.(2021b)]{2021ApJ...917...95Z} Zhang, L.-L., Ren, J., Huang, X.-L., et al.\ 2021b, \apj, 917, 95. \dodoi{10.3847/1538-4357/ac0c7f}

\end{thebibliography}
\end{document}